\documentclass[10pt, conference]{IEEEtran}
\usepackage{graphicx}
\usepackage{mhchem}
\usepackage{subfig}
\usepackage{url}
\usepackage{epstopdf}
\usepackage{amsmath}
\usepackage{amssymb}
\usepackage{balance}
\usepackage{cite}
\usepackage{citesort}
\usepackage[usenames, dvipsnames]{color}

\begin{document}
\bstctlcite{IEEEexample:BSTcontrol}

%0.67 (L) x 0.67 (R) x 0.81 (T) x 0.92 (B)

\title{Massive MIMO Performance Comparison of Beamforming and Multiplexing in the Terahertz Band}

\author{\IEEEauthorblockN{Sayed Amir Hoseini\IEEEauthorrefmark{1}\IEEEauthorrefmark{2}, Ming Ding\IEEEauthorrefmark{2} and Mahbub Hassan\IEEEauthorrefmark{1}\IEEEauthorrefmark{2}} 	\\
	{\IEEEauthorrefmark{1}School of Computer Science and Engineering, University of New South Wales, Sydney,  Australia} \\
	{\IEEEauthorrefmark{2}Data61, CSIRO, Sydney, Australia}\\
		Email: s.a.hoseini@unsw.edu.au, Ming.Ding@data61.csiro.au, mahbub.hassan@unsw.edu.au}

\maketitle

\begin{abstract}
In this paper, we compare the performance of two main MIMO techniques, beamforming and multiplexing, in the Terahertz (THz) band. The main problem with the THz band is its huge propagation loss, which is caused by the tremendous signal attenuation due to molecule absorption of the electro-magnetic wave. To overcome the path loss issue, massive MIMO has been suggested to be employed in the network and is expected to provide Tbps for a distance within a few meters. In this context, beamforming is studied recently as the main technique to take advantage of MIMO in THz and overcome the very high path loss with the assumption that the THz communication channel is Line-of-Sight (LoS) and there are not significant multipath rays. On the other hand, recent studies also showed that the well-known absorbed energy by molecules can be re-radiated immediately in the same frequency. Such re-radiated signal is correlated with the main signal and can provide rich scattering paths for the communication channel. This means that a significant MIMO multiplexing gain can be achieved even in a LoS scenario for the THz band. Our simulation results reveal a surprising observation that the MIMO multiplexing could be a better choice than the MIMO beamforming under certain conditions in THz communications. 

\end{abstract}

\section{ Introduction}\label{sec:introduction}
To respond to the huge increasing demand for the wireless data traffic, recently the terahertz (THz) band (0.1-10 THz) is envisioned to make Tbps wireless link feasible \cite{AKYILDIZ2016massive}. In spite of the wide unused bandwidth in this spectrum, the high propagation loss is the main issue of using such spectrum. Thus, the potential applications of the THz link are limited to short range communications such as nanosensors \cite{zarepour2017semon}, wireless on-chip communications and wireless personal area networks \cite{Akyildiz201416}. Moreover, part of the radio signal attenuation at the THz frequencies is due to molecular absorption which is frequency selective and increases the total loss to more than 200 dB for some frequencies at 10-meters distance. 

Basically, to overcome the very high path loss the transmit power could be largely increased. Unfortunately, this is not feasible with the current technology and it is limited to a few of mW \cite{Akyildiz201416}. Alternately, channel gain can be significantly improved by means of the multi-antenna beamforming technique. Indeed, Due to the very small footprint of a large number of antennas at the THz band, beamforming using very large scale Multiple Input Multiple Output (MIMO) systems has been considered in the field as a practical solution which can provide up to 55 dB channel gain at 1 THz \cite{AKYILDIZ2016massive}.

However, beamforming comes at the cost of system complexity and signaling overhead where the transmitter should receive the channel state information continuously and align the beam to the receiver. On the other hand, to achieve a significant MIMO beamforming gain in high frequency spectrum the beam would become very narrow which is sometimes described as a \emph{pencil beam}. This makes beamforming vulnerable to any transmitter/receiver mobility because it is difficult to perform beam re-alignment in a very short time interval.

Another approach to take advantage of MIMO is the MIMO multiplexing technique. While the beamforming technique strives to focus the transmission energy and achieve a large channel gain in a specific direction, the multiplexing technique builds it strength on creating parallel information channels. However, the multiplexing gain is significant only when there are enough non-negligible multipath signal components in a rich scattering environment. Because of the huge path loss, THz communication is usually assumed to be applied in as a Line-of-Sight (LoS) dominant channel and thus, the research focus has been on beamforming rather than multiplexing.

However, recent studies show that in the channel medium, molecules absorb and re-radiate the electromagnetic energy in THz band ~\cite{Kokkoniemi2015discuss,Jornet2014a,jornet2013fundamentals,Akyildiz201416}, which transforms the LoS channel into a rich-scattering environment. The re-radiation is usually considered as noise but the theorical model shows it is highly correlated to main signal \cite{jornet2013fundamentals}. In this paper, we will theoretically investigate the THz channel capacity for both cases of beamforming and multiplexing in a MIMO set-up. We find that the multiplexing technique can provide a considerable capacity gain in comparison with the beamforming technique on certain conditions. Also, in some other conditions where the beamforming yields a higher capacity, the multiplexing technique is still preferable choice due to its easier implementation. Note that in this work we assume a multiplexing technique using a blind precoding scheme without channel state information (CSI). In contrast, the beamforming technique always requires accurate CSI to smartly direct its energy in the spatial domain.

The rest of the paper is structured as follows.
In Section~\ref{sec:model}, 
we present the molecular absorption model for the calculation of channel transfer function,
Section~\ref{sec:analysis} analyzes the MIMO channel model considering the molecular re-radiation, 
followed by simulation results in Section~\ref{sec:simulSetup}.
Finally,
we conclude the paper in Section~\ref{sec:con}.

%%%%%%%%%%%%%%%%%%%%%%%%%%%%%%%%%%%%%%%%%%%%%%%%%%%%%%%%%%%%%%%%%%%%%%%%%
%%%%%%%%%%%%%%%%%%%%%%%%%%%%%%%%%%%%%%%%%%%%%%%%%%%%%%%%%%%%%%%%%%%%%%%%%
%%%%%%%%%%%%%%%%%%%%%%%%%%%%%%%%%%%%%%%%%%%%%%%%%%%%%%%%%%%%%%%%%%%%%%%
%%%%%%%%%%%%%%%%%%%%%%%%%%%%%%%%%%%%%%%%%%%%%%%%%%%%%%%%%%%%%%%%%%%%%%%%%

\section{Channel model and MIMO capacity} \label{sec:model}

The molecular absorption model defines how different species of molecules in a communication channel absorb energy from the electromagnetic signals and how they re-radiate them back to the environment.
This section first explains the concept of \emph{absorption coefficient} used to characterize the absorption capacity of a given molecule species,
followed by the attenuation and re-radiation models that are built upon this coefficient.

\subsection{Molecular absorption coefficient}

The \emph{medium absorption coefficient},
$k(f)$,
at frequency $f$ is a weighted sum of the molecular absorption coefficients in the medium~\cite{Jornet2014a}, which can be formulated as
\begin{equation}
k(f) = \sum_{i=1}^{N} m_i k_{i}(f),
\label{eq:Kf}
\end{equation}
where $k_{i}(f)$ is the molecular absorption coefficient of species $S_i$ on condition of temperature $\mathcal{T}$ and
and pressure $\mathcal{P}$.
$k_{i}(f)$ can be obtained from HITRAN~\cite{Rothman2012Database}.
In this work,
to get the values of $k(f)$,
we will use some predefined standard atmosphere conditions and their corresponding ratio of molecules in the air,
which are tabulated in~\cite{Rothman2012Database}.

\subsection{Attenuation of radio signal}

The attenuation of the radio signal at the THz frequencies is due to spreading and molecular absorption. In more detail,
the spreading attenuation is given by
\begin{equation}
 A_{\rm spread}(f, d)  =  \left( \frac{4 \pi f d}{c}\right) ^2,
 \label{eq:Atten_Spread}
\end{equation}
where $c$ is the speed of light.
The attenuation due to molecular absorption is characterized as
\begin{equation}
A_{\rm abs}(f,d) =  e^{k(f) \times d},
\label{eq:Atten_abs}
\end{equation}
where $k(f)$ is the absorption coefficient of the medium at frequency $f$.

Thus,
the line-of-sight (LoS) received power at the receiver becomes
\begin{eqnarray}
  \nonumber P_{{\rm r,LoS}}(f,d) &=& P_{t}(f)\times\left(\frac{c}{4\pi fd}\right)^{2}\times e^{-k(f)\times d}.
    \label{eq:Atten_pr1}
\end{eqnarray}

\subsection{Molecular re-radiation}
\label{sec:molecular-reradiation}
The existing molecules in communication medium will be excited by electromagnetic waves at specific frequencies.
The excitement is temporary and the vibrational-rotational energy level of molecules will come back to a steady state and the absorbed energy will be re-radiated in the same frequency.
These re-radiated waves are usually considered as noise in the literature \cite{Akyildiz201416}.
Molecular absorption is not white and its power spectral density (PSD) is not flat because of the different resonant frequencies of various species of molecules.
The PSD of the molecular absorption noise that affects the transmission of a signal, $S_{N_{\rm abs}}$,
is contributed by the atmospheric noise $S^B_{N}$ and the self-induced noise $S^X_{N}$ as addressed in~\cite{Jornet2014a}:
\begin{align}
&S_{N_{\rm abs}}(f,d)= S^B_{N}(f,d)+S^X_{N}(f,d), \label{eq:NoisePDS}\\
&S^B_{N}(f,d)=lim_{d\to\infty} (k_B T_0 (1-e^{-k(f) d})) \Big( \frac{c}{\sqrt{4\pi} f} \Big)^2, \label{eq:Noise01}\\
&S^X_{N}(f,d)=P_t(f)(1-e^{-k(f) d}) \Big(\frac{c}{{4\pi f d}}\Big)^2, \label{eq:Noise02}
\end{align}
where $k(f)$ is the absorption coefficient of the medium at frequency $f$,
$T_0$ is the reference temperature ($ 296K) $,
$k_B$ is the Boltzmann constant,
$P_t(f)$ is the power spectral density of the transmitted signal and $c$ is the speed of light.
The first term in \eqref{eq:NoisePDS},
which is called sky noise and defined in \eqref{eq:Noise01} is independent of the signal wave.
However,
the self-induced noise in \eqref{eq:Noise02} is highly correlated with the signal wave~\cite{jornet2013fundamentals},
and can be considered as a distorted copy of the signal wave.
Thus, equation~\eqref{eq:Noise02} can be revised as the received power of the re-radiated signal by molecules at the receiver by
\begin{equation}
 P_{\rm r,a}(f,d)=P_t(f)(1-e^{-k(f) d}) \Big(\frac{c}{{4\pi f d}}\Big)^2.
\label{eq:Atten_pr2}
\end{equation}

Since the phase of the re-radiated wave depends on the phase of molecular vibration,
which varies from molecules to molecules~\cite{barron2004molecular},
the received power in this case is affected by a large number of phase-independent re-radiated photons.
Thus,
we assume a uniformly distributed \emph{random} phase for the received signal, with its power given by~\eqref{eq:Atten_pr2}.

\subsection{Channel Transfer Function}

\label{sec:Hfunction}
The channel transfer function for a single LoS channel is given by
\begin{equation}
\begin{aligned}
\tilde{h}_{\rm LoS}(f,d) &=\sqrt{ \left( \frac{c}{4 \pi f d}\right) ^2 e^{-k(f) \times d}} \times e^ {j2\pi\frac{d}{\lambda}} \\
			   &= \left( \frac{c}{4 \pi f d}\right)  e^{-k(f) \times \frac{d}{2}} \times e^ {j2\pi\frac{d}{\lambda}}.
\label{eq:Hfunc_1}
\end{aligned}
\end{equation}

Then,
the partial channel transfer function resulted from the molecular absorption and excluding the LoS component can be represented by
\begin{equation}
\begin{aligned}
\tilde{h}_{\rm a}(f,d)&=\sqrt{(1-e^{-k(f) d}) \Big(\frac{c}{{4\pi f d}}\Big)^2}\times e^ {j2\pi\beta_{\rm random}}\\
 			 &=(1-e^{-k(f) d})^{\frac{1}{2}} \Big(\frac{c}{{4\pi f d}}\Big) \times e^ {j2\pi\beta_{\rm random}}.
\label{eq:Hfunc_2}
\end{aligned}
\end{equation}
Hence,
the total channel transfer function is the superposition of the partial channel transfer functions,
which is written as

\begin{equation}
\begin{aligned}
\tilde{h}(f,d) = \tilde{h}_{\rm LoS}(f,d) + \tilde{h}_{\rm a}(f,d),
\label{eq:HfuncM}
\end{aligned}
\end{equation}

\begin{eqnarray}
  \nonumber \tilde{h}(f,d)  \hspace{-0.2cm}&=&\hspace{-0.2cm} \left( \frac{c}{4 \pi f d}\right)  e^{-k(f) \times \frac{d}{2}} \times e^ {j2\pi\frac{d}{\lambda}} \\
    \hspace{-0.2cm}& &\hspace{-0.2cm} + (1-e^{-k(f) d})^{\frac{1}{2}} \Big(\frac{c}{{4\pi f d}}\Big) \times e^ {j2\pi\beta_{\rm random}}.
    \label{eq:Hfunc}
\end{eqnarray}

\subsection{MIMO channel model and capacity}

In this paper,
we consider a MIMO system that is consisted of $n_t$ transmitting antennas and $n_r$ receiving ones.
The received signal vector $y$ at $n_r$ receiving antennas can be formulated as
%~\cite{chinani2003MIMOcap}
\begin{equation}
y=\tilde{H}x+n,
\label{eq:MIMObase}
\end{equation}
where $x$ is the transmitted signal vector form $n_t$ transmitting antennas,
and $n$ is an $n_r\times1$ vector with zero-mean independent noises with variance $\sigma^2$.
$\tilde{H}$ is the channel matrix 

where each of its elements, ${\tilde{h}_{ij}}$, is a complex value denoting the transfer coefficient associated with the $j$th transmitter antenna and the $i$th receiver antenna.
Note that ${\tilde{h}_{ij}}$ can be obtained from \eqref{eq:Hfunc} for frequency $f$ and distance $d_{ij}$.

The capacity of MIMO channel can be written as
\begin{equation}
C = {\rm log}_{2}{\rm det} ({I}_{n_r}+\frac{P}{n_t\sigma^2}\mathbf{\tilde{H}}\mathbf{\tilde{H}}^\dagger),
\label{eq:MIMOcapBase}
\end{equation}
%where $\rho$ is the average signal-to-noise ratio (SNR) per receiving antenna
where $P$ is total transmitting power,
 and $I$ is the identity matrix 
 %\cite{chinani2003MIMOcap}.
Since the determinant of $({I}_{n_r}+\frac{P}{n_t\sigma^2}\mathbf{\tilde{H}}\mathbf{\tilde{H}}^\dagger)$ can be computed by the product of the eigenvalues of the matrix $\mathbf{\tilde{H}}\mathbf{\tilde{H}}^\dagger$,
%and the zero eigenvalues of matrix $\mathbf{H}\mathbf{H}^\dagger$ do not contribute to such product,
the MIMO capacity can thus be written in the form of a product of non-zero eigenvalues as~\cite{Tse2005MIMObook}
\begin{equation}
C = \sum_{i=1}^{\kappa}{\rm log}_2(1+\frac{P\lambda_i^2}{\kappa\sigma^2}),
\label{eq:MIMOcapEig}
\end{equation}
%where $\mathbf{H}$ is normalaized channel matrix,
where
$\lambda_i$ denotes singular values of the matrix $\mathbf{\tilde{H}}$,
and hence the squared singular values $\lambda_i^2$ denotes the eigenvalues of the matrix $\mathbf{\tilde{H}}\mathbf{\tilde{H}}^\dagger$. Each of the $\lambda_i^2$ characterize an equivalent information channel where $\frac{P\lambda_i^2}{k\sigma^2}$ is the corresponding signal-to-noise ratio (SNR) of the channel at the receiver.
Note that $\kappa$ denotes the number of non-zero $\lambda_i^2$,
which for beamforming technique it is equal to one and in multiplexing technique it could be the rank of $\mathbf{\tilde{H}}$ with $\kappa\leq {\rm min}(n_r,n_t)$~\cite{Tse2005MIMObook}. However, because we use blind precoding and uniform power allocation for multiplexing technique $\kappa=n_t$. Therefore, equation \eqref{eq:MIMOcapEig} is valid for uniform power allocation at the transmitter. Furthermore, the equivalent channel SNR, $\frac{P\lambda_i^2}{\kappa\sigma^2}$, should meet a minimum receiver threshold to be reliably detectable by the receiver. In this paper, we assumed 0 dB as the SNR threshold and uniform power allocation at the transmitter.

The main difference between beamforming and multiplexing techniques is how to tune or exploit the eigenvalue distribution. In more details, beamforming technique aims to maximize $\lambda_1$ to improve the channel SNR for a single data stream while in the multiplexing technique, a uniform eigenvalue distribution is preferable. In this way, multiplexing technique can utilize parallel data streams through MIMO and maximize the data rate. The complexity of beamforming comes from eigenvalues tuning because it means the channel state information (CSI) should be measured and sent back to the transmitter periodically for optimum precoding. This also results in a protocol overhead in the channel. On the other hand, multiplexing gain can take advantage of eigenvalue value distribution even with a blind precoding. This is more beneficial when there is a rich scattering environment in the channel. In next section, we will discuss how the re-radiation can provide a rich scattering environment.

%%%%%%%%%%%%%%%%%%%%%%%%%%%%%%%%%%%%%%%%%%%%%%%%%%%%%%%%%%%%%%%%%%%%%%%%%
%%%%%%%%%%%%%%%%%%%%%%%%%%%%%%%%%%%%%%%%%%%%%%%%%%%%%%%%%%%%%%%%%%%%%%%%%

\section{Analysis on the channel with molecular absorption}
\label{sec:analysis}

\begin{figure*}[t]
	\begin{center}
		\subfloat[K-factor]{\label{fig:k-factor}
\includegraphics[width=0.66\columnwidth ,clip=true, trim=0 0 0 0]{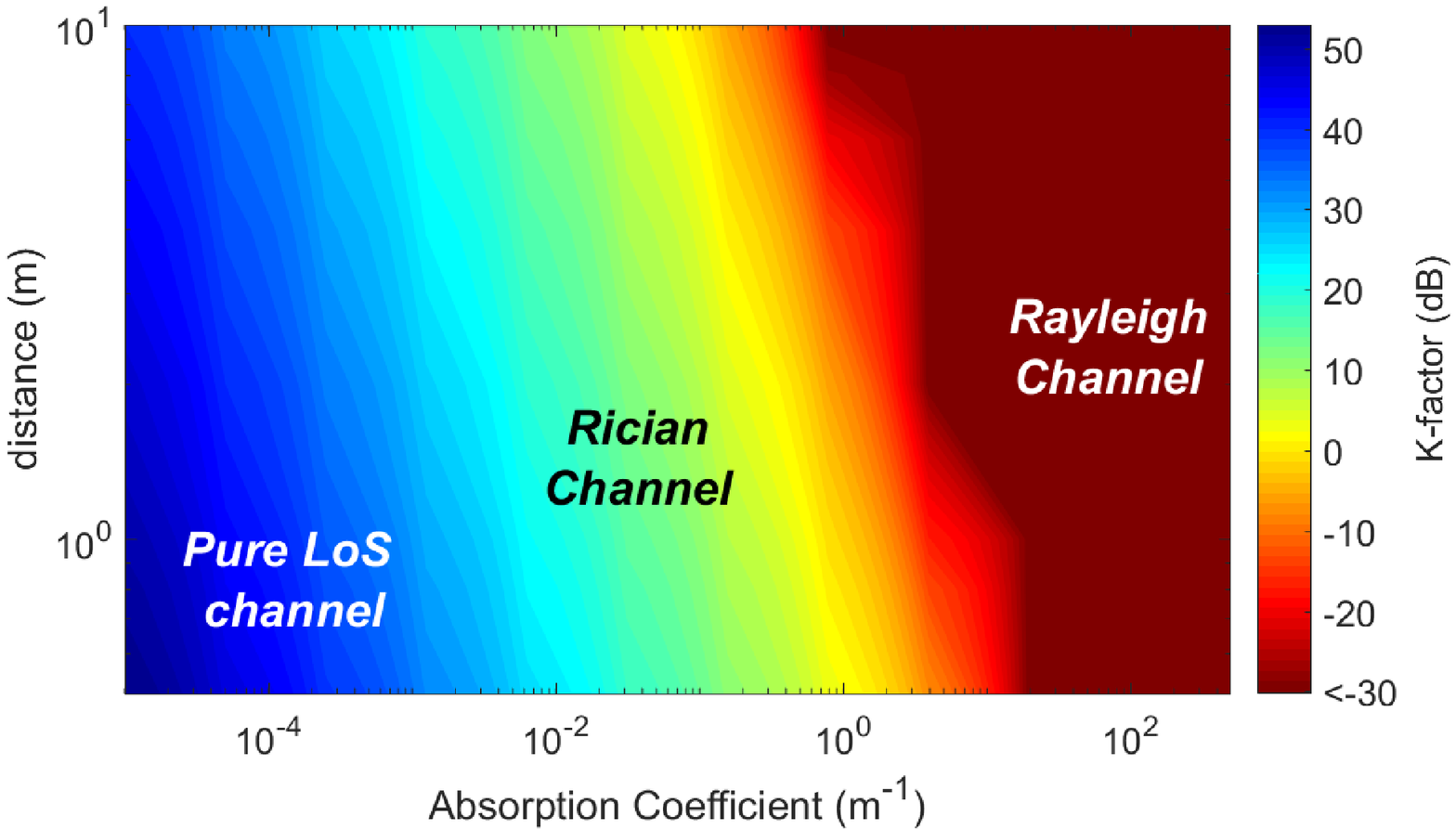}}
		\subfloat[MIMO capacity using beamforming]{\label{fig:k-factCapBeam}
\includegraphics[width=0.66\columnwidth ,clip=true, trim=0 0 0 0]{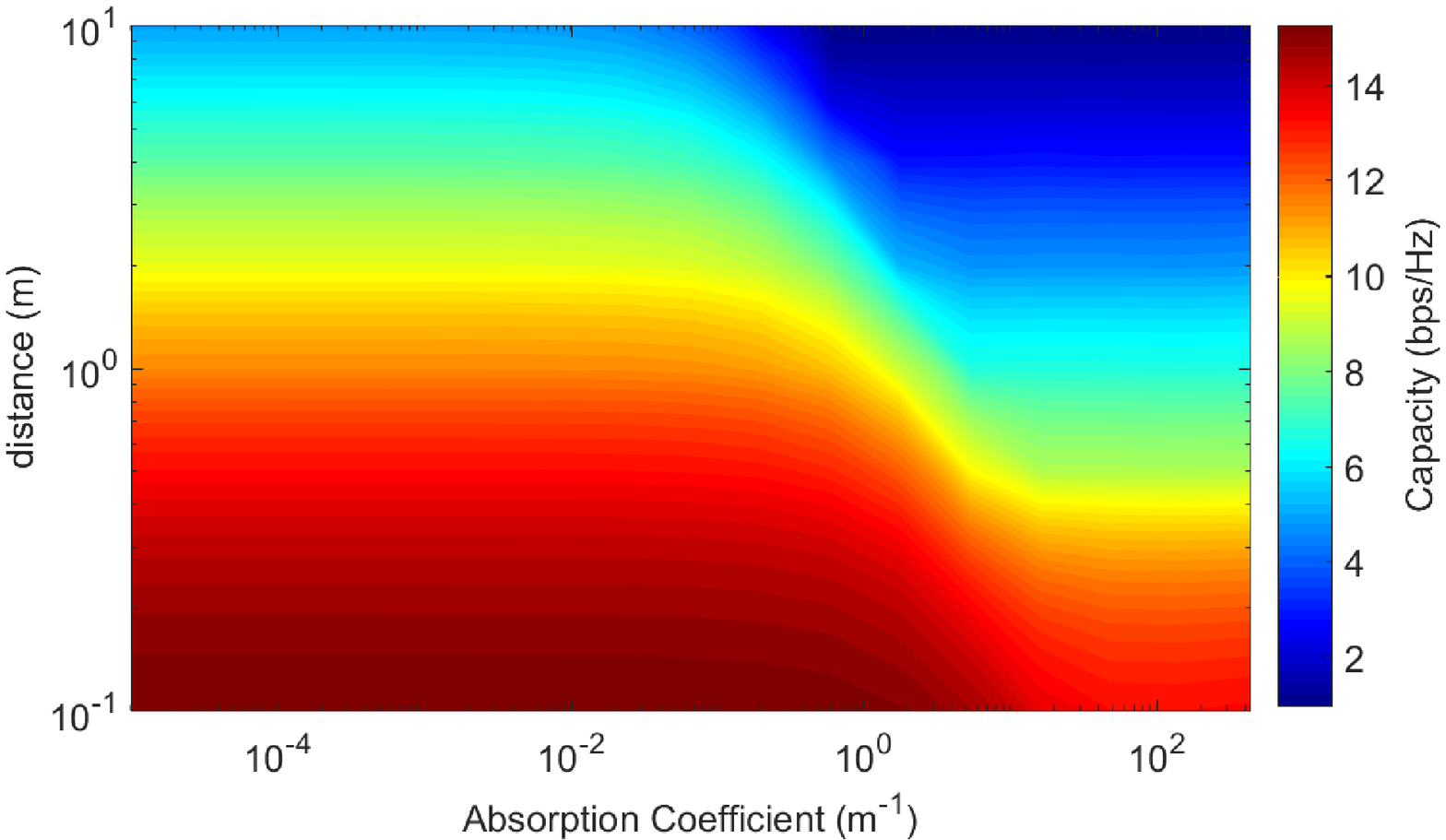}}
		\subfloat[MIMO capacity using multiplexing]{\label{fig:k-factCapMP}
\includegraphics[width=0.66\columnwidth ,clip=true, trim=0 0 0 0]{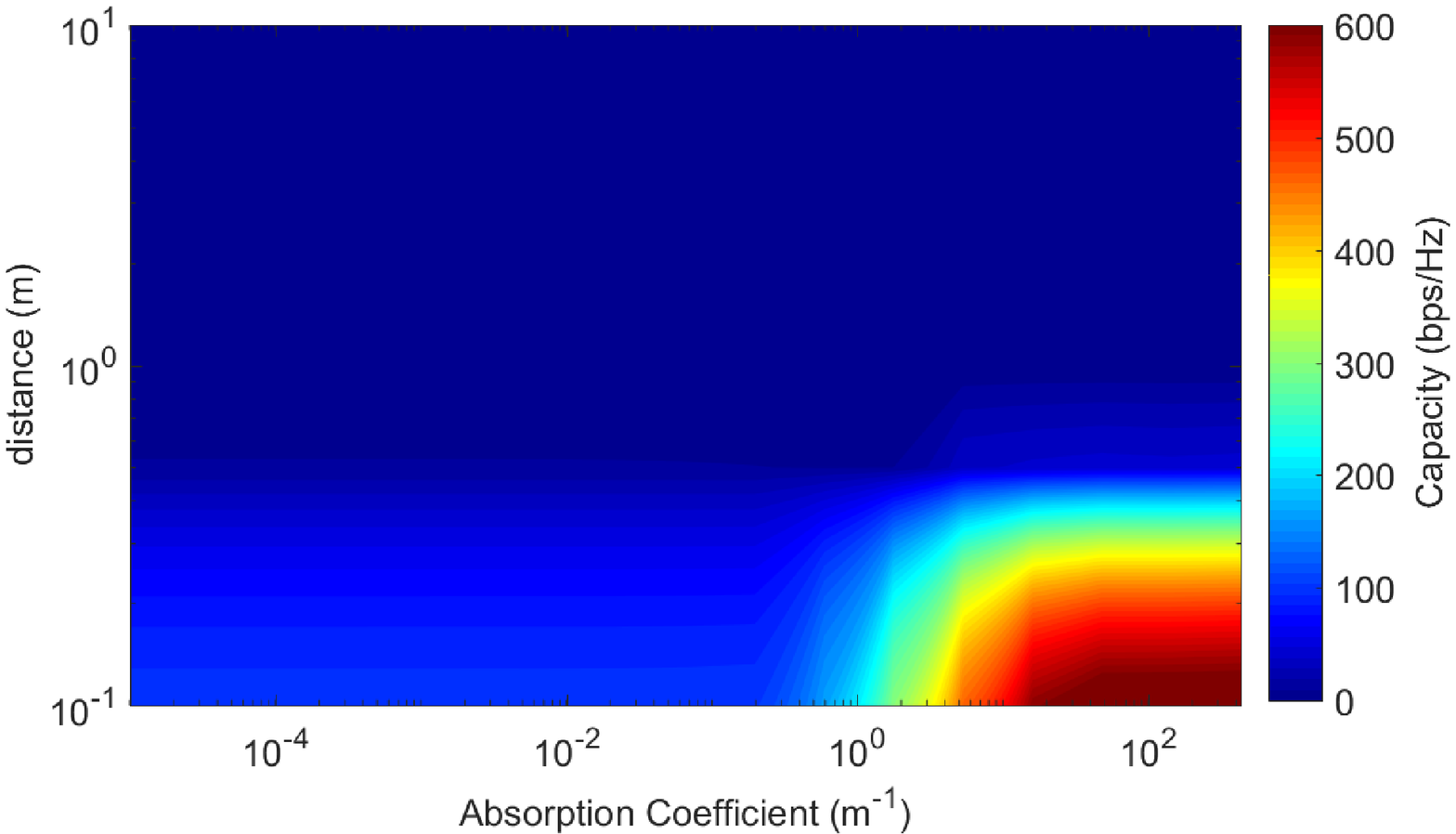}}
		\caption{K-factor is an increasing function of distance and absorption coefficient. For both multiplexing and beamforming techniques, the performance gain is affected by K-factor. The capacity is calculated for 225x225 MIMO system.}	
		\label{fig:K-factor_analys}
	\end{center}
\vspace{-0.5cm}
\end{figure*}

To analyze the MIMO channel capacity and characterize the scattering richness of channel quantitatively, lets decompose and normalize channel transfer function $\tilde{H}$ as

\begin{equation}
\begin{aligned}
{H}(f,d) = \sqrt{\frac{K}{K+1}}{H}_{\rm LoS}(f,d) + \sqrt{\frac{1}{K+1}}{H}_{\rm a}(f,d),
\label{eq:HfuncM2}
\end{aligned}
\end{equation}

where $H$, ${H}_{\rm LoS}$ and ${H}_{\rm a}$ are normalized with corresponding channel gain. Because of uniformly distributed random phase of received re-radiated signal,  elements of ${H}_{\rm a}$ are independent and identically
distributed (i.i.d) complex Gaussian random variables with zero
mean and unit magnitude variance. $K$ is the ratio of powers of the LoS signal and the re-radiated components and if we assume the channel distance is much longer than antenna space, it can be obtained by

\begin{equation}
\begin{aligned}
K = \frac{P_{\rm r,LoS}(f,d)}{P_{\rm r,a}(f,d)} = \frac{e^{-k(f) d}}{1-e^{-k(f) d}}.
\label{eq:k-factor}
\end{aligned}
\end{equation}

This is same as the well-known Rician channel model where the $K$ is called \textit{Rician K-factor}. Equivalently, K-factor shows how much channel is rich in term of scattering and multi-path rays. Equation \eqref{eq:k-factor} shows $K$ is a function of absorption coefficient of channel medium $k(f)$ and the distance between transmitter and receiver $d$ so that a longer distance and a higher absorption result smaller $K$, as shown in Figure \ref{fig:k-factor}. The capacity of MIMO channel considering Rician K-factor is studied in several works 
%\cite{Jayaweera2002Rician}, 
\cite{farrokhi2001link,lebrun2006mimo}. Authors in \cite{lebrun2006mimo} showed the lower bound of Rician channel expected capacity for large number of antennas is the expected capacity of channel considering only NLoS component,

\begin{equation}
\begin{aligned}
E(C(H))\geq E(C(\sqrt{\frac{1}{K+1}}H_a)),
\label{eq:lower-band1}
\end{aligned}
\end{equation}

\begin{equation}
\begin{aligned}
 \implies E(C(H))\geq E(C(\sqrt{1-e^{-k(f) d}}H_a)),
\label{eq:lower-band2}
\end{aligned}
\end{equation}

where $E(.)$ denotes the expectation. It is clear that the lower band is a increasing function of absorption coefficient, $\forall f_1 , f_2 \quad \text{such that} \quad k(f_2)\geq k(f_1),\quad  E_{\substack{min}}(C(f_2)) \geq E_{\substack{min}}(C(f_1))$.

%%%%%%%%%%%%%%%%%%%%%%%%%%%%%%%%%%%%%%%%%%%%%%%%%%%%%%%%%%%%%%%%%%%%%%%%%
%%%%%%%%%%%%%%%%%%%%%%%%%%%%%%%%%%%%%%%%%%%%%%%%%%%%%%%%%%%%%%%%%%%%%%%%%
\section{Simulation and discussion}
\label{sec:simulSetup}

\begin{table}[b]
\centering
\caption{Simulation parameters}
\label{tab:simParam}
\begin{tabular}{|l|l|}
\hline
Transmitter and receiver distance ($d$) & $0.1, 1, 10$ m                       \\
Inter-element spacing ($s$)             & $0.5 \lambda$ (wave length)        \\
Transmitter arrays angle ($\phi$)          & $90^\circ $ \\
Receiver arrays angle ($\theta$)                   & $90^ \circ$ \\
Number of arrays on each side ($n$)     & $225$ \\
Transmit power & $0, 10$ dBm \\ 
Noise power & $-80$ dBm \\ \hline
%Receiver antenna gain & $10$ dBi \\ 
\end{tabular}
\vspace{-0.5cm}
\end{table}

\begin{table*}[t]
\centering
\caption{Atmosphere standard gas mixture ratio in percentage for different climates \cite{Rothman2012Database}}
\scriptsize
\label{tab:gasMix}
\begin{tabular}{|l|l|}
\hline
USA model, mean latitude, summer, H=0 & H2O: 1.860000   CO2: 0.033000   O3: 0.000003   N2O: 0.000032   CO: 0.000015   CH4: 0.000170   O2: 20.900001   N2: 77.206000 \\ \hline
USA model, mean latitude, winter, H=0 & H2O: 0.432000   CO2: 0.033000   O3: 0.000003   N2O: 0.000032   CO: 0.000015   CH4: 0.000170   O2: 20.900001   N2: 78.634779 \\ \hline
USA model, high latitude, summer, H=0 & H2O: 1.190000   CO2: 0.033000   O3: 0.000002   N2O: 0.000031   CO: 0.000015   CH4: 0.000170   O2: 20.900001   N2: 77.876781 \\ \hline
USA model, high latitude, winter, H=0 & H2O: 0.141000   CO2: 0.033000   O3: 0.000002   N2O: 0.000032   CO: 0.000015   CH4: 0.000170   O2: 20.900001   N2: 78.925780 \\ \hline
USA model, tropics, H=0               & H2O: 2.590000   CO2: 0.033000   O3: 0.000003   N2O: 0.000032   CO: 0.000015   CH4: 0.000170   O2: 20.900001   N2: 76.476779 \\ \hline
\end{tabular}
\vspace{-0.5cm}
\end{table*}

\subsection{Simulation set-up}
\label{sec:geometry}

\begin{figure*}[]
    \begin{center}
            \subfloat[absorption coefficient, T= $273~K$, P= $1~atm$]{\label{fig:kflog}
\includegraphics[width=0.75\columnwidth ,clip=true, trim=0 0 0 0]{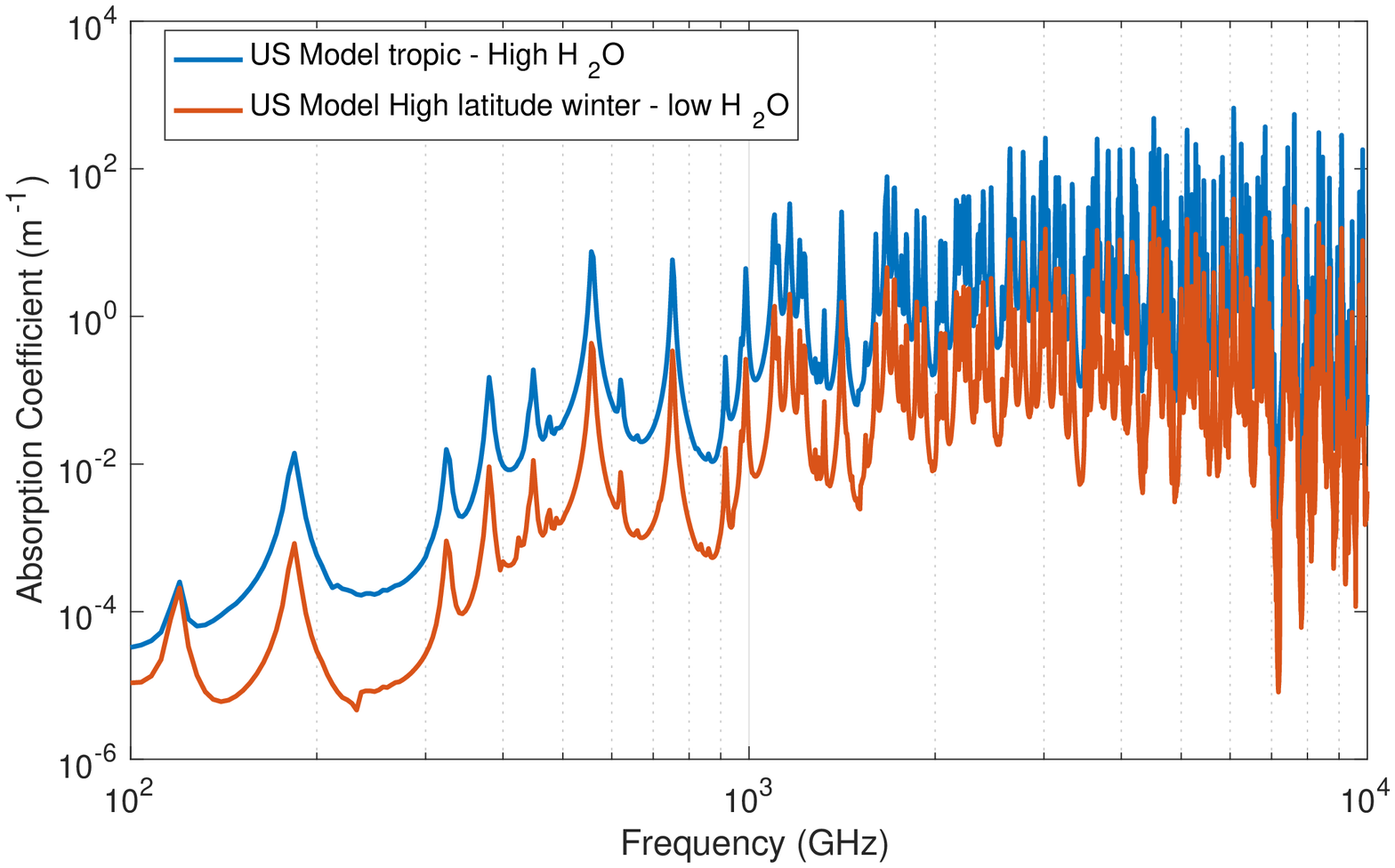}} 
			\subfloat[Signal Attenuation]{\label{fig:att}
\includegraphics[width=0.75\columnwidth ,clip=true, trim=0 0 0 0]{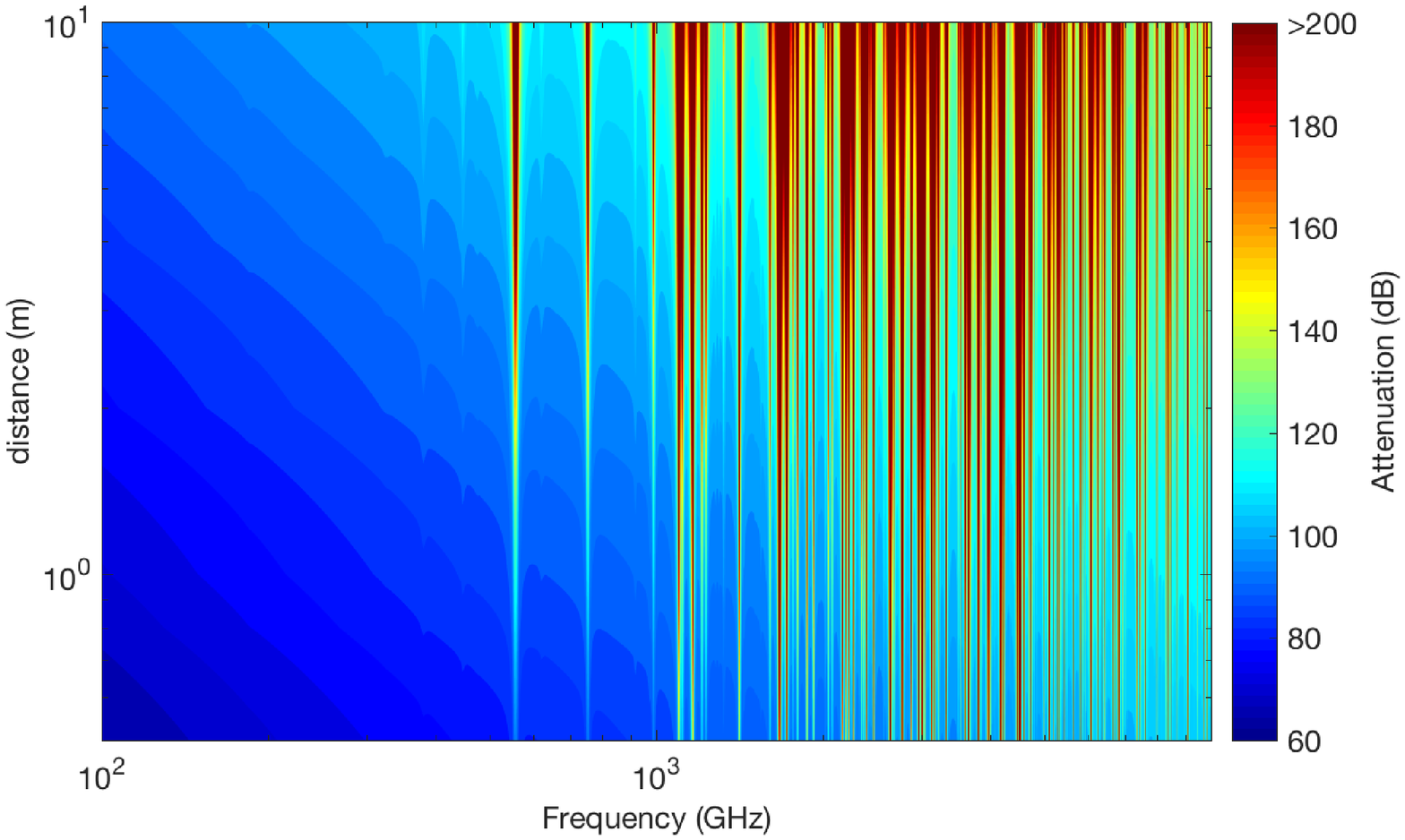}} \vspace{-0.3cm} \\
            \subfloat[distance=0.1m, transmit power=1mW]{\label{fig:cam0-1m1mW}
\includegraphics[width=0.75\columnwidth ,clip=true, trim=0 0 0 0]{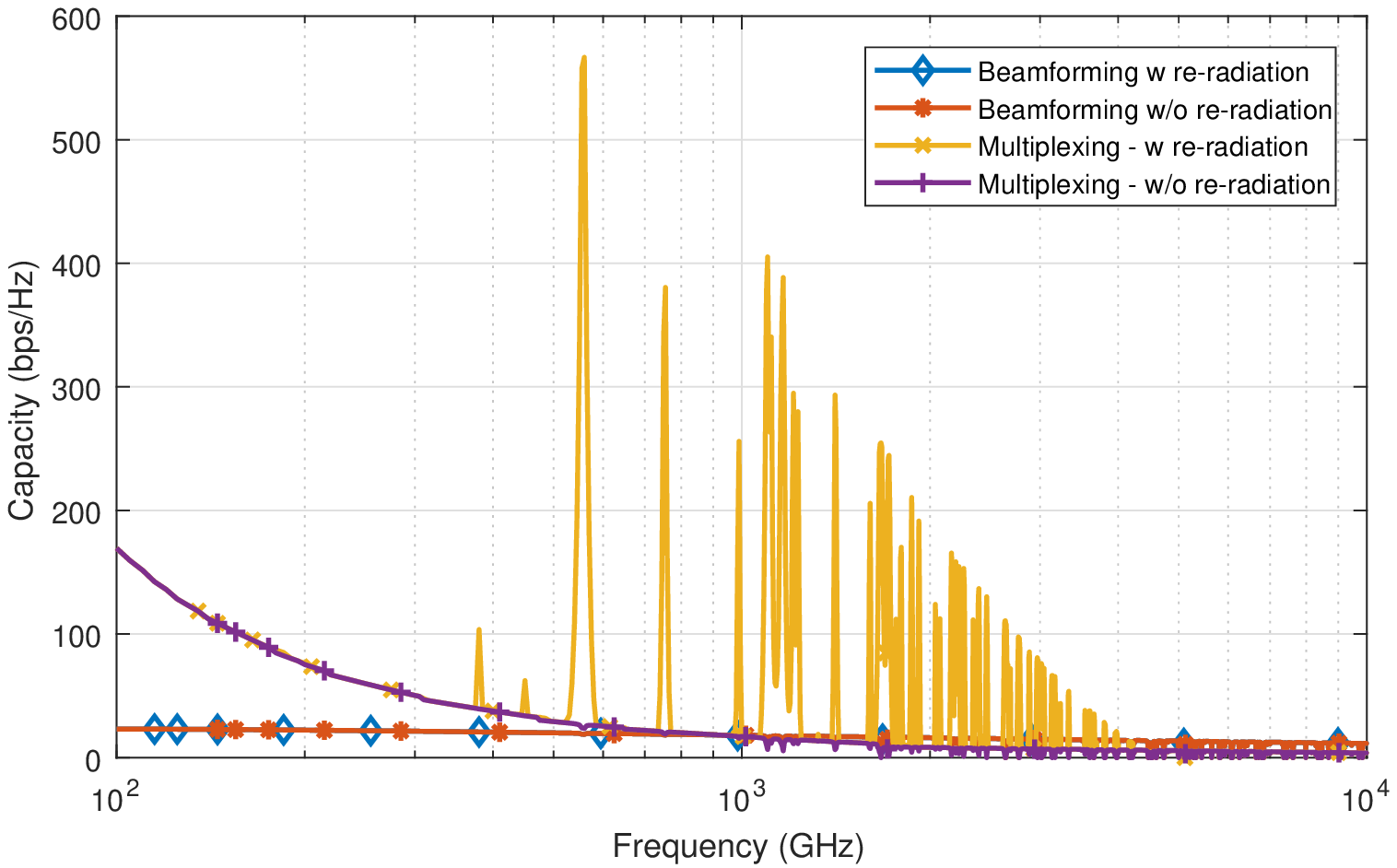}}
        \subfloat[distance=0.1m, transmit power=10mW]{\label{fig:cam0-1m10mW}
\includegraphics[width=0.75\columnwidth ,clip=true, trim=0 0 0 0]{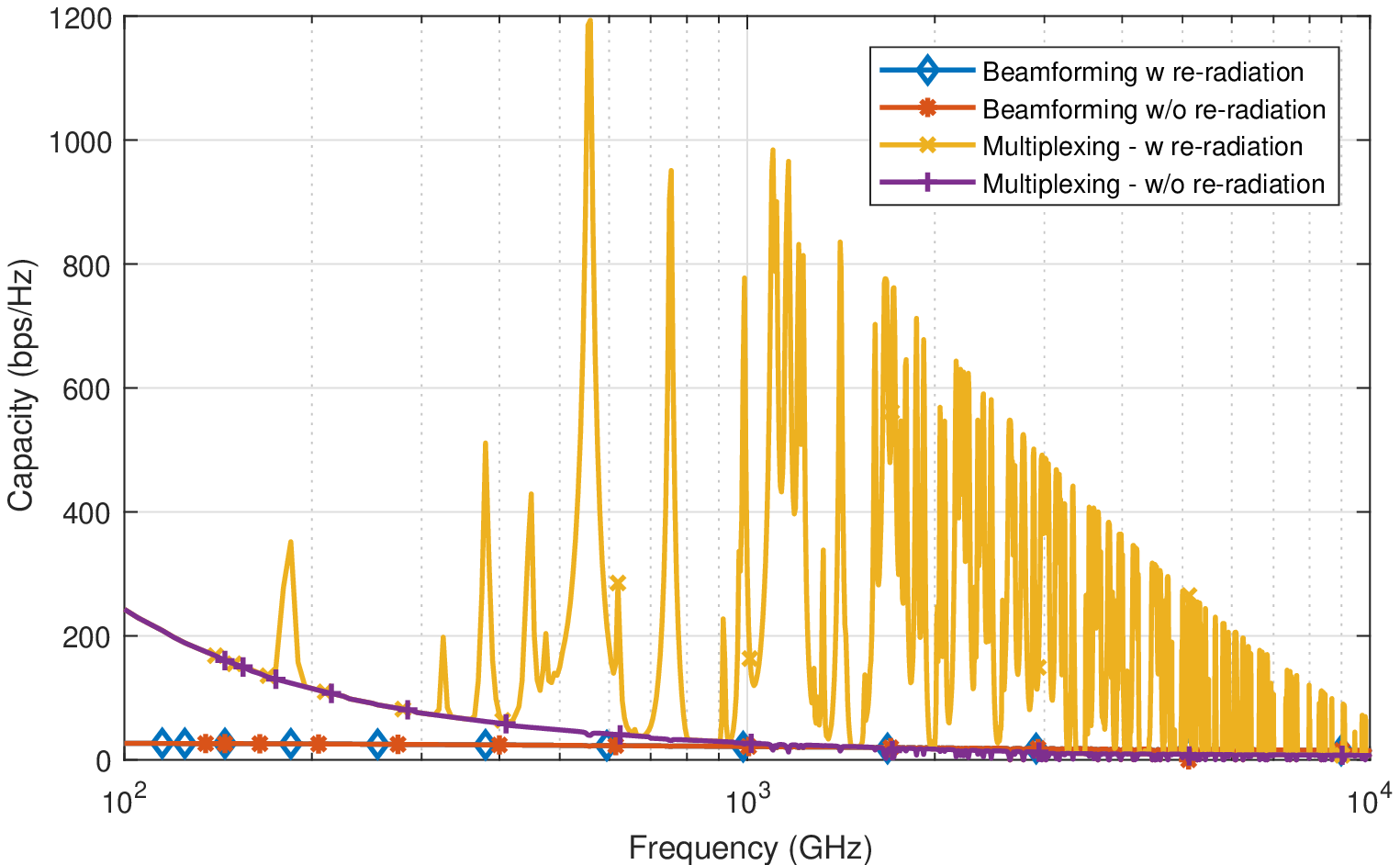}}\vspace{-0.2cm} \\
        \subfloat[distance=1m, transmit power=1mW]{\label{fig:cam1m1mW}
\includegraphics[width=0.75\columnwidth ,clip=true, trim=0 0 0 0]{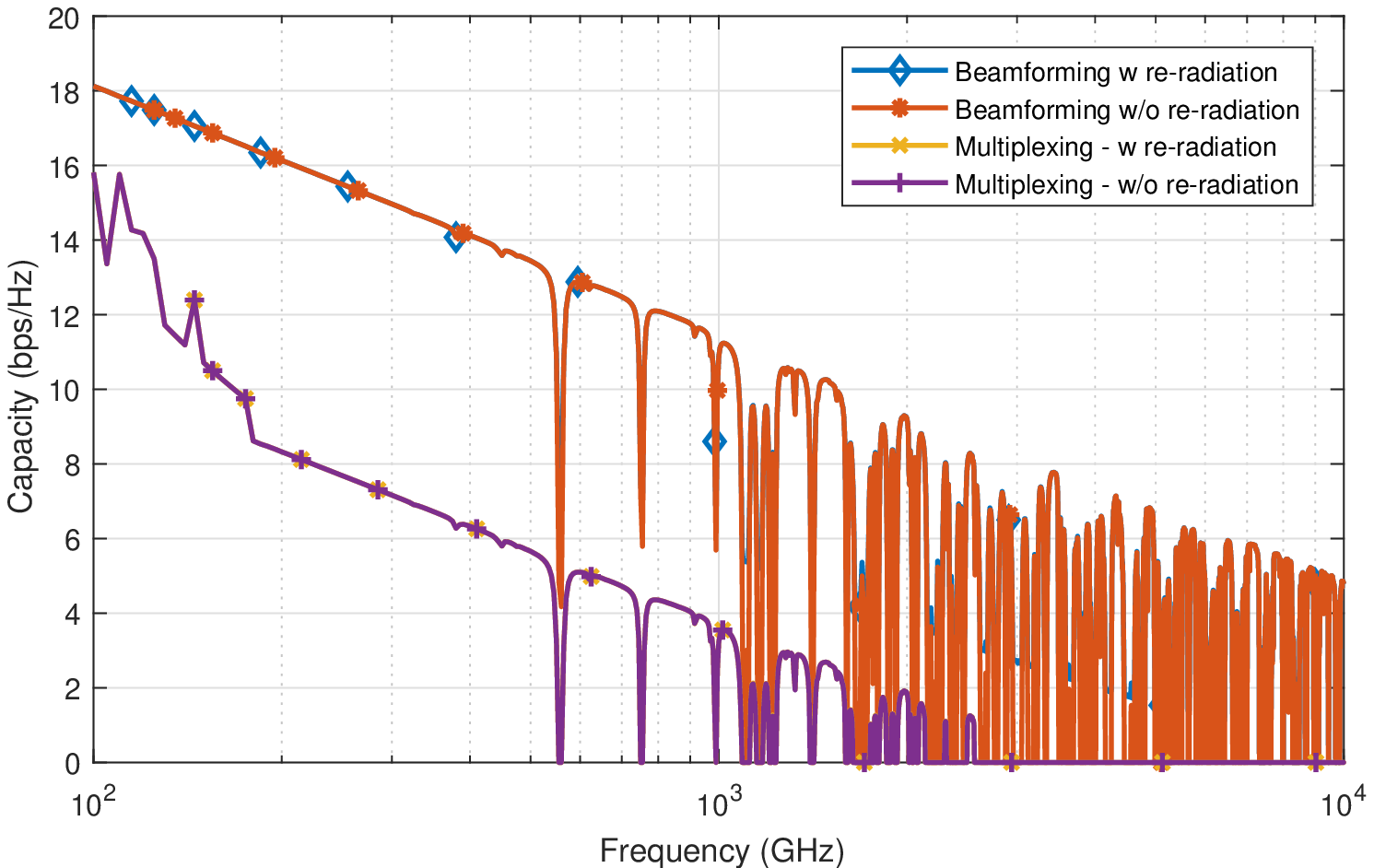}}
        \subfloat[distance=1m, transmit power=10mW]{\label{fig:cam1m10mW}\includegraphics[width=0.75\columnwidth]{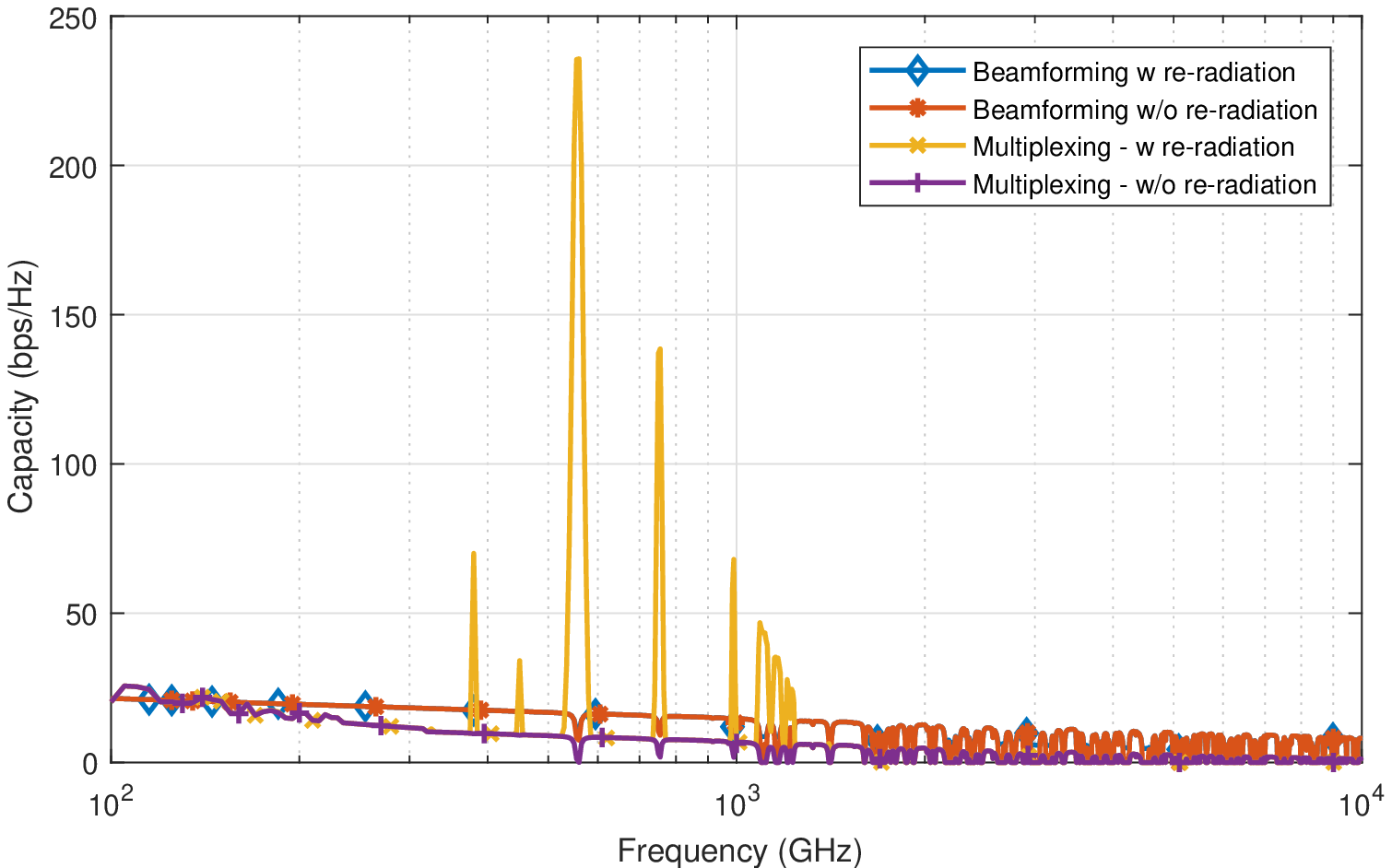}}\vspace{-0.2cm}\\
        \subfloat[distance=10m, transmit power=1mW]{\label{fig:cam10m1mW}
\includegraphics[width=0.75\columnwidth ,clip=true, trim=0 0 0 0]{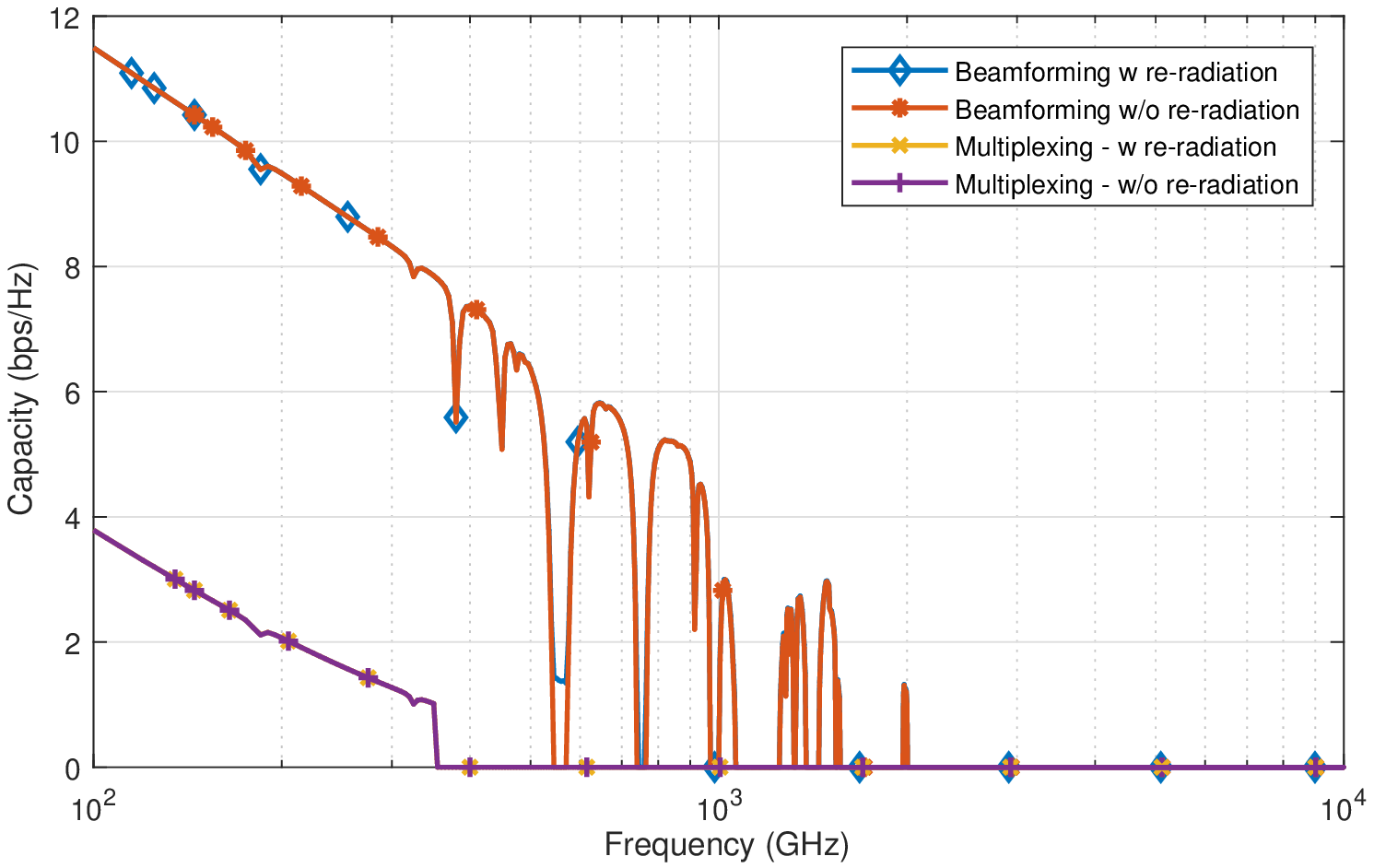}}
        \subfloat[distance=10m, transmit power=10mW]{\label{fig:cam10m10mW}\includegraphics[width=0.75\columnwidth]{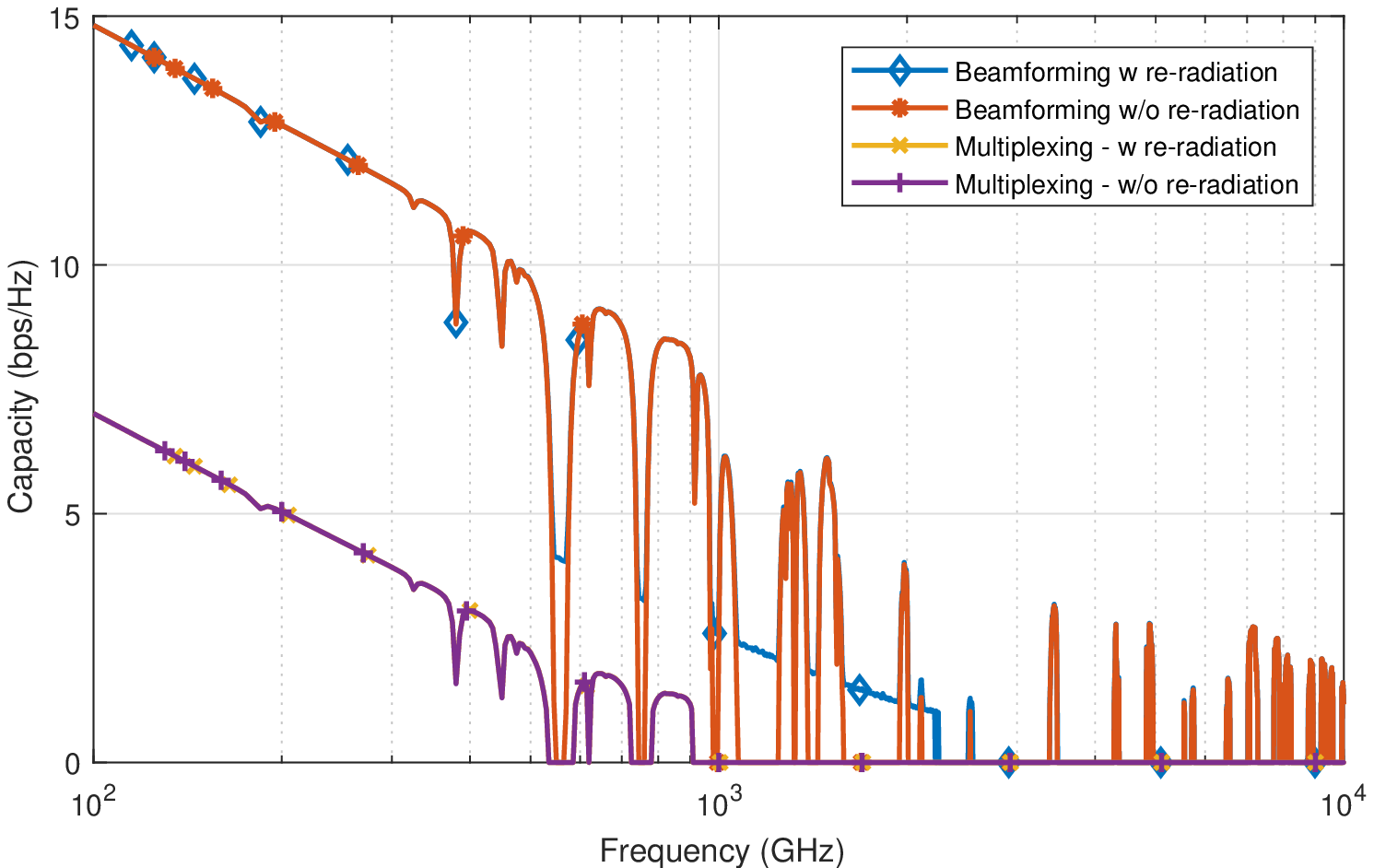}}            
        \caption{225x225 MIMO channel performance in tropic atmosphere.}    
        \label{fig:cap1}
    \end{center}
\vspace{-0.5cm}
\end{figure*}

In this section, to evaluate the molecular absorption impact on THz MIMO capacity,
%To evaluate the MIMO capacity in THz and show the performance impact of the molecular absorption,
we consider a simple $n\times n$ MIMO system with a square uniform Arrays,
where at both transmitter and receiver, the inter-element spacing $s$ is equal to half of the wavelength
and the channel distance is $d$.
%The considered MIMO system is illustrated in Figure~\ref{fig:geometry}.
Moreover,
we consider uniform power allocation to transmitter arrays operating in an open-space LoS scenario.
The default values of the parameters are listed in Table~\ref{tab:simParam},
and different values will be explained when necessary. Since we apply random phases on NLoS components created by molecular re-radiation,
we conduct the evaluation of the MIMO capacity with molecular re-radiation for 1000 times and show the average result.

We use the online browsing and plotting tools\footnote{http://hitran.iao.ru/gasmixture/simlaunch},
which is based on HITRAN databases \cite{Rothman2012Database} to generate absorption coefficients for different single gas or some predefined standard gas mixture of the atmosphere at sea level,
as shown in Table \ref{tab:gasMix}.
Since the water molecules play main roles in a normal air environment at THz bands
we use the highest and lowest water ratio in Table~\ref{tab:gasMix},
i.e., the \emph{"USA model, high latitude, winter"} and \emph{"USA model, tropics"}.
The corresponding absorption coefficients in THz bands have been shown in Figure~\ref{fig:kflog} for an ambient temperature of 273\,K and a sea level pressure of 1\,atm. For a tropic atmosphere,
  the water ratio is higher than that of the winter atmosphere,
  and thus we can see a significant increase in the absorption coefficient among these two gas mixtures.

In our simulation,
we assume a constant transmit power over the entire frequency spectrum
and display the MIMO capacity in bps/Hz for THz bands. We consider a MIMO set-up with 225 antennas at each side in a uniform square planar array. Our aim is to compare the beamforming and multiplexing techniques in different channel conditions. First, we calculate the channel capacity for beamforming while the re-radiation is totally ignored in the channel. Next, the beamforming capacity is re-calculated when the re-radiation is taken into account. Finally, the multiplexing gain is calculated with and without the consideration of re-radiation. In all scenarios, capacity is obtained by~\ref{eq:MIMOcapEig}.

In the first step, the simulation is run at 500 GHz with the practical range of absorption coefficient (${10^{-5}\sim10^{+3}}$) over the THz spectrum, as shown in Figure \ref{fig:K-factor_analys}. It should be noted that the actual value of absorption coefficient at 500 GHz is shown in Figure \ref{fig:kflog}. The beamforming and multiplexing techniques capacity is calculated for a range of 0.1$\sim$10 m distance and a 1 mW transmit power.

Secondly, the channel is simulated for two different transmit power and three distances with realistic absorption coefficients. Our assumption on the transmit power is based on current technology \cite{Akyildiz201416} and a previous work on THz massive MIMO \cite{AKYILDIZ2016massive}. Furthermore, distances have been chosen to cover various application scenarios. For example, THz nanosensors are considered to communicate in a very short distance in the order of 0.1-10 cm or less, while THz communications are also nominated to provide terabit per second ultra high video communication link at around 1 m distance for home entrainment devices like TV or virtual reality (VR). In addition, longer distances to a few meters characterize wireless personal or local networks. Simulation results are presented in Figure \ref{fig:cap1}.

\subsection{ The MIMO Capacity vs. the K-factor}
Figure \ref{fig:K-factor_analys} illustrates how the channel is transformed from a LoS dominant channel to a Rayleigh channel and how it effects on the MIMO beamforming and multiplexing capacity gain. As can be seen in Figure \ref{fig:k-factCapBeam}, the beamforming gain is decreasing when the absorption coefficient increases which is because in the very high absorption, the channel is not LoS dominant anymore and there is significant NLoS signal component generated by molecule re-radiation or equivalently lower K-factor. In contrast, Figure \ref{fig:k-factCapMP} shows the multiplexing technique takes advantage of higher absorption to reach a huge data rate. However, the low SNR limit the multiplexing gain in longer distances so that it drops sharply to zero beyond 2 m. In Figure \ref{fig:cap1}, more results for the THz spectrum with realistic absorption coefficients will be presented.

\subsection{ The MIMO Capacity vs. the Transmit Power and Distance}\label{sec:simulResult}

The channel attenuation including molecular attenuation in \eqref{eq:Atten_abs} and spreading attenuation in  \eqref{eq:Atten_Spread} is illustrated in Figure \ref{fig:att}. While the spreading attenuation is increasing linearly in dB with distance and frequency, the molecular attenuation is also increasing with distance but is frequency selective. For example, while the total loss at 10m is 107 dB for 500 GHz, the total attenuation at 550 GHz is 86 dB at 1 m and it grows to 220 dB at 10 m which is mostly because of very high absorption of water molecules in the channel medium at this frequency. Note that the channel atmosphere for this case is from tropic data where the ratio of water molecules in the air is more than 0.02, as shown in Table \ref{tab:gasMix}.

Figure \ref{fig:cam0-1m1mW} and \ref{fig:cam0-1m10mW} illustrate the capacity of the investigated transmission techniques for a 10 cm distance. The transmit power is increased from 1 mW in Figure \ref{fig:cam0-1m1mW} to 10 mW in Figure \ref{fig:cam0-1m10mW}. It can be seen that a huge performance difference exists between multiplexing and beamforming, thanks to the tremendous multiplexing gain provided by the rich scattering environment due to molecule re-radiation. Furthermore, in very high absorption frequencies which existing studies consider as infeasible windows  for THz communications, a significant capacity improvement can be observed. This is because more absorption leads to more re-radiation, which transforms a LoS dominant channel to Rayleigh channel. The details can be found in Section \ref{sec:analysis}, where we have discussed about how the re-radiation decreases the K-factor and creates a rich scattering environment. To sum up, the re-radiation improves the multiplexing gain which is fundamentally supported by a better eigenvalue distribution and channel matrix rank in mathematical analysis.

In Figure \ref{fig:cam1m1mW} and \ref{fig:cam1m10mW}, the distance is increased to 1 m. With a relatively large distance for THz communications, it can be seen the beamforming gain is comparable with the multiplexing gain. However, we can see the multiplexing gain in high absorption windows, such as 540-580 GHz, is significantly higher than the rest of spectrum for a 10 mW transmit power. It is a different story for a 1 mW transmit power where the capacity drops to zero in high absorption windows because the equivalent SNR, \large ($\frac{P\lambda_i^2}{k\sigma^2}$)\normalsize, of most parallel channels created by the multiplexing technique is less than 0~dB and practically such parallel channels are useless because the receiver can not reliably detect the received signals. Such results are not surprising since it has been shown in several works on conventional communication band \cite{gesbert2003theory} that the multiplexing performance drops dramatically in low SNR. However, considering the implementation challenges of beamforming, the multiplexing technique might still be a preferable choice for frequency up to 1 THz. For example, it can be observed in Figure \ref{fig:cam1m1mW} at 0.9 THz, the capacity is 4 and 11.7 bps/Hz for the multiplexing and beamforming techniques, respectively. 

Finally, Figures \ref{fig:cam10m1mW} and \ref{fig:cam10m10mW} present the results for a 10 m distance. For such a distance, path loss leads to a very low reception SNR and thus the beamforming performance is significantly better than the multiplexing performance. It is well-known that beamforming technique is not very effective where there are strong multipath rays \cite{gesbert2003theory}. Thus, it is observed that in very high absorption frequency windows, the beamforming performance drops sharply. It is not only because of receiving strong NLoS rays caused by molecule re-radiation but also due to LoS signal attenuation. Note that the multiplexing technique can take advantage of same windows in high SNR as we discussed above for Figure \ref{fig:cam1m10mW}.

\balance
%\vspace{0.1cm}
\section{Conclusion}
\label{sec:con}
In this paper, we compared the beam forming and multiplexing techniques of MIMO in the terahertz band. We showed in high SNR, high transmit power or lower distance, the multiplexing technique can provide a considerable capacity gain compared with beamforming. However, for beyond a few meters such as 10 meters, there should be enough transmitting power possibility to use multiplexing technique, otherwise the capacity drops to zero where the beamforming technique can still provide effective spectrum efficiency at the cost of complexity and protocol overhead. Our theoretical model also showed re-radiation of molecules in the THz band can be helpful for massive MIMO system to improve the channel performance using multiplexing technique. The re-radiation can provide significantly strong multipath components to achieve a full spatial multiplexing gain where the receiver is in an enough SNR coverage. It means some very high absorption frequency windows which have been formerly pointed as not feasible for communication might be more preferable choices for MIMO in some certain applications.  
%\vspace{0.1cm}
\bibliographystyle{IEEEtran}
\bibliography{Ref}
\end{document}